\documentclass[prl,twocolumn,preprintnumbers,amsmath,amssymb,superscriptaddress]{revtex4}
\usepackage{color}
\usepackage{graphicx}

\begin{document}
\title{Femtosecond study of the interplay between excitons, trions, and carriers in (Cd,Mn)Te quantum wells}
\author{P.~P\l ochocka} \affiliation{Institute of Experimental
Physics, Warsaw University, Ho\.za 69, 00-681 Warsaw, Poland}
\author{P.~Kossacki}
\affiliation{Institute of Experimental Physics, Warsaw University,
Ho\.za 69, 00-681 Warsaw, Poland}
 \author{W.~Ma\'slana}
 \affiliation{Institute of Experimental Physics, Warsaw University,
Ho\.za 69, 00-681 Warsaw, Poland}
 \affiliation{Laboratoire de
Spectrom\'etrie Physique, CNRS et Universit\'e Joseph
Fourier-Grenoble, B.P.87, 38402 Saint Martin d'H\`{e}res Cedex,
France}
\author{J.~Cibert}
\author{S.~Tatarenko}
\affiliation{Laboratoire de Spectrom\'etrie Physique, CNRS et
Universit\'e Joseph Fourier-Grenoble, B.P.87, 38402 Saint Martin
d'H\`{e}res Cedex, France}
\author{C.~Radzewicz}
\author{J.~A.~Gaj}
\email{Jan.Gaj@fuw.edu.pl}
\affiliation{Institute of Experimental
Physics, Warsaw University, Ho\.za 69, 00-681 Warsaw, Poland}
\date{\today}
\begin{abstract}
We present an absorption study of the neutral and positively
charged exciton (trion) under the influence of a femtosecond,
circularly polarized, resonant pump pulse. Three populations are
involved: free holes, excitons, and trions, all exhibiting
transient spin polarization. In particular, a polarization of the
hole gas is created by the formation of trions. The evolution of
these populations is studied, including the spin flip and trion
formation processes. The contributions of several mechanisms to
intensity changes are evaluated, including phase space filling and
spin-dependent screening. We propose a new explanation of the
oscillator strength stealing phenomena observed in p-doped quantum
wells, based on the screening of neutral excitons by charge
carriers. We have also found that binding heavy holes into charged
excitons excludes them from the interaction with the rest of the
system, so that oscillator strength stealing is partially blocked.
\end{abstract}
\maketitle

Since the first experimental observation of charged excitons
(trions) in semiconductor quantum wells (QW) \cite{Kheng93} it has been
commonly recognized that optical spectra of moderately doped
QWs contain  neutral- and charged exciton lines, whose
parameters are influenced by the presence of free carriers.
Amongst the effects of carriers on the oscillator strength of the
exciton, phase space filling (PSF) and carrier-carrier interaction
are usually distinguished \cite{Schmitt-Rink89}. PSF
arises due to carriers with the same spin as that of the
carrier entering the exciton; thus, in the case of photocreated
carriers, the pump and probe beams have to be co-polarized.
Carrier - carrier interaction (screening) takes place between
carriers of any spin. However, as pointed out in
\cite{Schmitt-Rink89}, a cut-off of the screening is expected at
the closest approach of like particles, as given by the size of
the exchange-correlation hole. This effect, due to the Pauli
exclusion principle, makes the screening by carriers of opposite
spins more efficient.  This applies to the neutral exciton
 when the free carrier  and the carrier within the
exciton have opposite spins (e.g., with photocreated carriers when
the pump and probe beams are cross polarized), but not to the
trion, which contains one carrier of each spin. In the same
spirit, diffusion of the exciton on carriers was considered in
\cite{Lejeune98} to be responsible for the polarization-selective
broadening of the exciton line due to the presence of a polarized
electron gas in GaAs-based quantum wells. A decrease of the
exciton oscillator strength is systematically observed
\cite{Kheng93,Koss03,Brink99} in transmission spectra of
CdTe-based QWs in the presence of carriers with opposite spins
(either electrons or holes). As it is the presence of these
carriers which makes the formation of trions possible, this effect
produces a balance between the decrease of the neutral exciton
intensity and the increase of the trion intensity, which is often
referred to as "oscillator strength stealing" (OSS). Theoretical
efforts have been undertaken to describe OSS
\cite{Suris01,Esser01} but no commonly accepted model has been
devised to establish the existence of a "sum rule" between the
intensities of the neutral and charged excitons with the same
polarization. In what follows we reexamine the origin of
OSS thanks to a sub-picosecond
pump-probe experiment on a (Cd,Mn)Te/(Cd,Mg)Te QW. We first
show that under our experimental conditions, PSF plays a minor
role in the variation of the oscillator strength of (neutral) excitons,
thus confirming a previous observation from CW experiments on (Cd,Mn)Te
QWs where the hole gas was spin polarised thanks to the giant Zeeman
effect \cite{Koss99}. Then we show that spin-dependent carrier-carrier
interaction is at the origin of the OSS. To this purpose, we use a
pump-probe transmission experiment where we establish a controlled
population of free carriers, neutral
 excitons, and charged excitons. Actually, we demonstrate a clear
mechanism for the creation of a spin polarization in a 2D hole
gas. Previously reported studies of the creation of a spin
polarization in a 2D carrier gas were focused on n-type doping,
and obtained thanks to the fast relaxation of the holes within the
photocreated electron-hole pair\cite{Croo96}. Here we show that the formation
of charged excitons is an efficient way of inducing a polarization
of the free carriers.

\begin{figure}[t]
\begin{center}
\includegraphics[width=.35\textwidth]{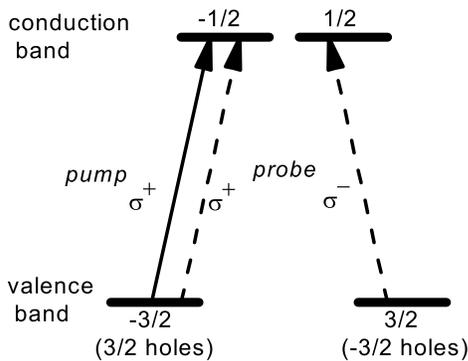}%
\end{center}
\caption{Optical transitions for pump and probe
pulses}\label{Fig0}
\end{figure}

The time resolved study was carried out on a modulation doped
structure consisting of a single $80$ \AA~quantum well (QW) of
Cd$_{1-x}$Mn$_{x}$Te $($x $\approx 0.0018)$ embedded between
Cd$_{0.66}$Zn$_{0.07}$Mg$_{0.27}$Te barriers grown
pseudomorphically on a $($100$)$ Cd$_{0.88}$Zn$_{0.12}$Te
substrate. Due to strain and confinement, only heavy holes have to
be considered. Modulation p-type doping was assured by a
nitrogen-doped layer at $200$ \AA~from the QW. The density of the
hole gas in the QW was controlled by an additional illumination
with photon energy above the gap of the barriers, provided by a
tungsten halogen lamp with a blue filter: the mechanism and its
calibration are described in detail in \cite{Koss99}. The sample
was mounted strain-free in a superconducting magnet and immersed
in superfluid helium at $1.8$ K. The pulses were generated by a
Ti$^{3+}: $Al$_{2}$O$_{3}$ laser tuned at $765$ nm ($1620$ meV),
at a repetition rate of $100$ MHz. The $100$ fs duration of the
laser pulse assured a spectral width of about $40$ nm (80 meV),
much broader than the splitting between the neutral and the
charged excitons. The pump and probe pulses were focused on the
sample to a common spot of diameter smaller than $100$ $\mu$m, and
the spectrum of the probe pulse transmitted through the sample was
recorded as a function of the pump-probe delay. The power of both
pulses was controlled independently, the pump-to-probe intensity
ratio being at least $20:1$. The average power of the pump beam
was typically $16$ mW, which results in the creation of a few
times $10^{10}$cm$^{-2}$ excitons by each pulse, depending on the
experimental conditions. No biexciton signature was observed in
any of our experiments. The pump beam was polarized circularly (by
convention, $\sigma^{+}$). It created electrons with spin $-1/2$
in the conduction band, and holes of momentum $+3/2$ in the  -3/2
spin-down valence subband, as shown on Fig. \ref{Fig0}. In some
experiments selective, resonant excitation of neutral or charged
excitons was achieved by shaping the laser pulse to a spectral
width less than 1 nm and a duration of about 2 ps. Thus the pump
pulse created either neutral excitons (X), or charged excitons
(X$^{+}$) involving in the latter case the binding of an
additional spin-down hole of momentum -3/2. The probe beam was
detected behind the sample at both circular polarizations, thus
measuring the absorption associated with the creation of either X
or
 X$^{+}$ having electron-hole pairs of the same
spin $(\sigma^{+})$ as the pump, or opposite $(\sigma^{-})$.

\begin{figure}[t]
\begin{center}
\includegraphics[width=.48\textwidth]{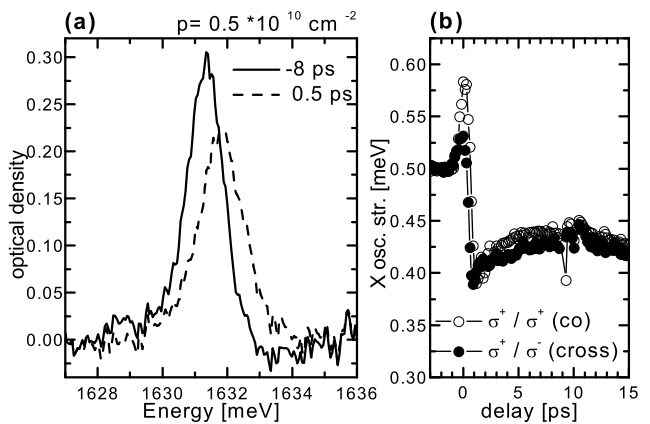}%
\end{center}
\caption{$($a$)$ Optical density for empty quantum well for
negative delay ($-8 $ps, solid line) and very short positive
 delay ($0.5$ ps, dashed line), for
co-polarized beams; $($b$)$ neutral exciton oscillator strength vs
delay. Open circles denote co-polarized beams, closed circles
cross-polarized. } \label{Fig1}
\end{figure}

Fig. \ref{Fig1}a shows typical transmission spectra before and
shortly after the pump pulse, measured on the QW with a negligible
 carrier density: Only the neutral exciton line is
observed. The pump pulse induces a broadening of the exciton line,
but also a decrease of its intensity, as measured using a Gaussian
fit of the line. This is shown in Fig.\ref{Fig1}b as a function of
the pump-probe delay, in both circular polarizations. At very
short delays we observe a peak due to coherent nonlinear effects,
useful for fixing the zero on the time scale. In what follows we
focus on the phenomena following this ultrashort transient. The
observed intensity reduction is similar for both polarizations.
This suggests that the main effect is the (spin-independent)
screening of the electron-hole interaction by the photo-created
excitons. This is in contrast with observations reported for III-V
QWs, where a much stronger effect for co-polarized beams has been
interpreted in terms of PSF \cite{Miller99}. Hence, screening of
the exciton by the carriers of photocreated excitons appears to
dominate over PSF in such a $($Cd,Mn$)$Te QW. A theoretical
estimate of the effect of PSF \cite{Schmitt-Rink85}, predicts it
to be proportional to the area occupied by the exciton in the
quantum well. Substituting the parameters of our QW we obtain a
decrease of the oscillator strength by a few percent. This is much
smaller than the experimental value (about 20 percent), supplying
an additional argument for a minor role of PSF.

\begin{figure*}[bt!]
\begin{center}
\includegraphics[width=.75\textwidth]{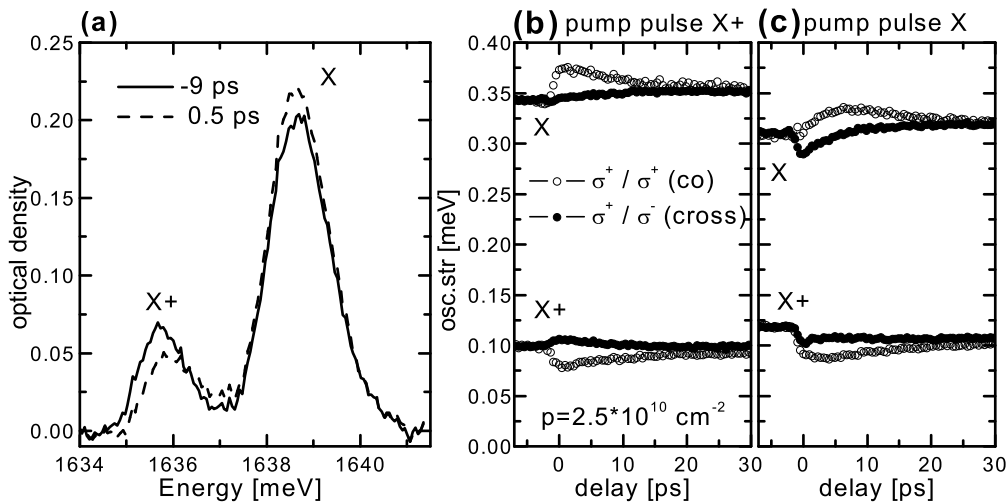}
\end{center}
\caption{\label{Fig2} $($a$)$Optical density for QW with hole
concentration p $\approx 2.5\times10^{10}$cm $^{-2}$. Solid line
denotes negative delays $(-8 $ps$)$, dash line is for short
positive delays ($0.5$ps)
 for co polarized when the pump pulse was tuned on X$^+$ absorption.
$($b$)$ evolution of neutral X and charged X$^+$ exciton
oscillator strength if pump pulse is tuned to charged exciton.
$($c$)$ evolution of X and X$^+$ if pump pulse is tuned to neutral
exciton.}
\end{figure*}

\begin{figure}[b]
\begin{center}
\includegraphics[width=.4\textwidth]{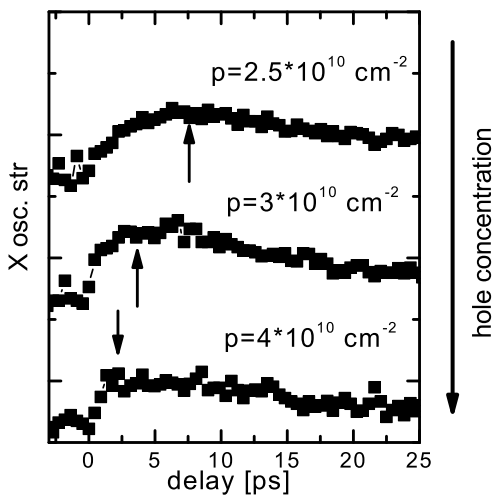}
\end{center}
\caption{\label{Fig3} Evolution of the X oscillator strength for different hole
densities, as indicated, with the pump beam tuned to X.
line.}
\end{figure}

We turn now to the case of the QW with carriers, so that both the
neutral exciton and the trion lines are visible. On transmission
 spectra before and shortly after the pump pulse (Fig.
\ref{Fig2}a), we observe a decrease of the trion intensity and an
increase of the neutral exciton intensity. The results shown in
Fig. \ref{Fig2}b, c, obtained from Gaussian functions fitted to
transmission spectra, will be explained below assuming that PSF is
negligible, and that screening by free carriers is more efficient
than that by carriers engaged in excitons $($neutral or
charged$)$. We first focus on the most pronounced effect, which is
observed in the co-polarized configuration with the pump beam
tuned to the trion line $($Fig.\ref{Fig2}b$)$. We observe a
decrease of the trion intensity and an increase of the neutral
exciton one. I.e., photocreating $(\sigma^{+})$ trions (excitons
bound to spin-down holes) has the same effect as decreasing the
density of the same spin-down holes in CW experiments where line
intensities have been studied as a function of background carrier
density \cite{Koss99}. We propose a common description of the
effects observed in the present pump probe experiment and those of
the CW experiments. Then we observed a linear dependence of the
trion oscillator strength on the population of holes with the
appropriate spin, at constant total hole density. In the
pump-probe experiment, a significant amount of the free carriers
with spin down become bound into trions. They are no more
available to form new trions in $\sigma^{+}$, so that the $X^+$
intensity decreases in this polarization. For the neutral exciton,
in CW experiments, a linear decrease was observed and attributed
to the OSS. We propose to interpret this OSS as the spin-dependent
part of screening of the neutral exciton by free holes. As
mentioned above, due to the Pauli exclusion principle, this
reduction is stronger in $\sigma^{+}$ polarization for free holes
with spin down. In the pump-probe experiment, those holes are used
to form the trions and therefore are excluded from the exchange
interaction with neutral excitons, so that the screening of X is
reduced. Our interpretation removes the conceptual difficulty of
the sum rule, which can be invoked to explain the cw results by
arguing that introduction of holes into the quantum well creates
new trion states at the expense of neutral exciton states. In our
pump-probe experiments the number of the trion states does not
change, only their occupation is modified by the pump pulse,
nevertheless the OSS effect occurs. The proposed interpretation
explains coherently both types of experiments.
 Note that when the
pump beam is tuned to the neutral exciton line, Fig.\ref{Fig2}c,
similar effects are observed, with a delay. The rise time of X
decreases with the initial hole density $($Fig.\ref{Fig3}$)$. We
identify this risetime with the trion formation time, estimated
using suitable rate equations (not shown) as 5 ps, 2 ps, and 1 ps
for hole densities $2.5 \times 10^{10} cm^{-2}$, $3 \times 10^{10}
cm^{-2}$, and $4 \times 10^{10} cm^{-2}$ respectively. As
expected, these values are shorter than the value of 65 ps
reported before \cite{Vanelle00,Koss03} for similar QWs with
smaller hole densities (the X and X$^+$ lines were of similar
intensities in \cite{Vanelle00,Koss03}, while in the present
sample the X line is absent in CW photoluminescence spectra
excited nonresonantly, not shown, proving a much higher hole
density). Besides, a higher temperature due to strong excitation
may contribute to the faster trion formation.

It is interesting to discuss now the weaker effects observed in the
cross-polarized configuration. When pumping into the $(\sigma^{+})$
neutral exciton, we create spin-up holes, thus increasing
the screening of the $(\sigma^{-})$ charged and neutral excitons. This
results in an intensity decrease of both the neutral and charged
excitons, which is small due to the binding of these holes within
excitons, and even weaker for the charged exciton due to the Pauli
exclusion principle. This is observed in Fig.\ref{Fig2}c. When pumping into
the trion however, in addition
to the former mechanism of screening by spin-up photocreated holes, we
trap free holes of spin-down into trions, thus reducing their efficiency
in the screening of the excitons. In the case of the neutral exciton, we
thus have a balance between two weak effects of opposite sign (spin-up
holes because they are bound in excitons and spin-down holes due to the Pauli exclusion
principle), so that a very small effect, if any, is seen. In the case
of the trion, screening by the photo-created spin-down holes is even further reduced
 by Pauli exclusion, so that one observes a fast increase of the
 $(\sigma^{-})$ trion intensity when pumping into the trion, due to the reduction of
screening by the spin-down free holes.

At longer pump-probe delays we observe a slow recovery of the
oscillator strength, which we attribute to spin relaxation. At
time delays longer than the trion formation time and smaller than
its recombination time, no matter whether the pump beam created
excitons or trions, we deal with two populations: trions and free
holes. The total population of these two species is constant, but
both exhibit a spin polarization. In particular, a nonequilibrium
spin polarisation of the free hole gas has been created by optical
excitation of the trions, which trap spin-down holes. Both
polarisation relax, and two spin flip processes have to be
considered: (i) spin flip of the free holes from the spin-up free
hole gas unaffected by the pump to the spin-down hole gas depleted
by the formation of $X^+$; (ii) spin flip of the $X^+$, which
actually is that of the electron bound in the trion. However, as
previously, we expect that the dominant contribution to screening
comes from the free holes. Thus, fitting the difference between
the exciton intensities measured in the two polarizations with an
exponential function, we determine the spin relaxation time of
free heavy holes, equal to 8 ps, weakly dependent on the hole
density. These values are comparable to the spin relaxation times
determined for holes bound in exciton and charged exciton
complexes. The photoluminescence experiments with much lower
excitation power and very similar heterostructures gave values
from about 3ps \cite{Cami01} through 20 ps \cite{Vanelle00} up to
35ps \cite{Koss03}. The difference
 might be a result of different Mn content, temperatures and density of
photo-created carriers. In particular, increase of k-vector and
scattering processes lead to shortening of spin flip time, as
predicted theoretically \cite{Bastard91}. It was shown
experimentally for a similar n-type sample \cite{Koss03} that for
the hole in X$^-$ the spin relaxation time may be decreased from
35ps to below 5ps by varying the excitation energy.
To conclude, the evolution of the charged and neutral exciton line
intensities (including oscillator strength stealing) is well
explained by neglecting phase space filling and assuming that: -
the trion intensity increases with the population of free carriers
with the opposite spin - neutral exciton screening is spin
dependent, trion is not, in accordance with the Pauli exclusion
principle - screening by free carriers is more efficient than
screening by carriers engaged in excitons.

This work has been partially supported by KBN grants
PBZ-KBN-044/P03/2001 and 5P03B02320, and Polonium program.

\bibliography{Femto}
\end{document}